\documentclass[twocolumn,aps,pre,amsmath]{revtex4}
\usepackage{graphicx}
\usepackage{dcolumn}
\usepackage{bm}
\usepackage{amssymb}
\usepackage{amsmath}
\usepackage{bm}
\usepackage{color}
\usepackage{bbold}
\usepackage{verbatim}
\usepackage{pstricks}
\usepackage{booktabs}
\usepackage{pst-plot}
\usepackage{amsfonts}

\def\be{\begin{equation}}
\def\ee{\end{equation}}

\begin{document}

\title{Effects of shear flow on phase nucleation and crystallization}

\author{Federica Mura}
\affiliation{Department of Physics, Ludwig-Maximilians-University Munich, Theresienstrasse 37, 80333 Munich, Germany}
\author{Alessio Zaccone}
\affiliation{Statistical Physics Group, Department of Chemical Engineering and Biotechnology, University of Cambridge, New Museums Site, Pembroke Street, CB2 3RA Cambridge, U.K.}
\date{\today}

\date{\today}
\begin{abstract}
Classical nucleation theory offers a good framework for understanding the common features of new phase formation processes in metastable homogeneous media at rest. However, nucleation processes in liquids are ubiquitously affected by hydrodynamic flow, and there is no satisfactory understanding of whether shear promotes or slows down the nucleation process. We developed a classical nucleation theory for sheared fluids systems starting from the molecular-level of the Becker-Doering master kinetic equation, and analytically derived a closed-form expression for the nucleation rate. 
The theory accounts for the effect of flow-mediated transport of molecules to the nucleus of the new phase, as well as for the mechanical deformation imparted to the nucleus by the flow field. 
The competition between flow-induced molecular transport, which accelerates nucleation, and flow-induced nucleus straining, which lowers the nucleation rate by increasing the nucleation energy barrier, gives rise to a marked non-monotonic dependence of the nucleation rate on the shear-rate. The theory predicts an optimal shear-rate at which the nucleation rate is one order of magnitude larger than in the absence of flow.
\end{abstract}

\maketitle

\section{Introduction}
Understanding the mechanism of shear-induced nucleation processes~\cite{Debenedetti,Greer} could play an essential role in life sciences where different phenomena, like protein~\cite{Penkova} and peptide~\cite{Forsyth} aggregation and crystallization, commonly take place under applied shear flows. For example, under in vitro conditions, shear is ubiquitous due e.g. to stirring of the solution. Under \textit{in vivo} and physiological conditions, protein aggregation, condensation and crystallization phenomena occur under cytoplasmic flow conditions~\cite{Goldstein}.

Anomalies in \textit{in vivo} protein crystallization are responsible for different pathological conditions. For example, the crystallization of the mutated hemoglobin inside human blood cells underlies numerous condensation diseases leading to anemia~\cite{Lawrence}.
Furthermore, the cytoplasmic flows inside embryos may drive P granules condensation during the specification of germ cells, a process not fully understood in which flow-enhanced nucleation could play an important role~\cite{Brangwynne}.

In a very different setting, shear-induced crystallization in the supercooled melt is of vital importance in metallurgy~\cite{Greer}. In particular, understanding the effect of shear on crystallization rate is crucial in the processing of metallic glasses which are cooled very rapidly from the high-temperature melt. For example, recent experiments reported a significant acceleration of the crystallization rate in supercooled metallic melts~\cite{Osuji}. 
Finally, crystallization under shear, in spite of being poorly understood, is a critical process in many industrial applications where shear flow is ubiquitous in continuous industrial processing and devices~\cite{langer}. Here we can just recall the pervasive role of shear flow in the industrial crystallization of pharmaceutical molecules~\cite{baird}. In the integrated modelling of industrial processes there is considerable need of analytical models which incorporate the basic microscopic molecular physics of the system, in terms of molecular interaction parameters, solvent properties, etc. 

Due to the pivotal role of nucleation processes in many fields, several experiments and simulations have been performed in an attempt to rationalize the effect of shear on nucleation. 
Very different outcomes have been reported with different materials and in different conditions range. In particular, while some studies have reported that shear flow essentially slows down the nucleation rate~\cite{Okubo,Blaak}, other studies have found that shear flow significantly boosts or accelerates the nucleation rate~\cite{Harrowell, Holmqvist, Coccorullo, Forsyth, Penkova, Mackley, Kerrache}. Pioneering simulations on colloids with Yukawa (screened-Coulomb) repulsion showed that the nucleation barrier increases quadratically with the shear rate~\cite{Blaak}, but the overall effect of shear on the nucleation rate was not reported. More recent simulations~\cite{Mokshin,Cerdà,Speck} suggested the possibility that a maximum in the nucleation rate versus shear rate may appear. 

Kinetic models for nucleation in shear flow have been proposed, for example using mesoscopic nonequilibrium thermodynamics~\cite{Rubi}. The latter study leads to a mesoscopic Smoluchowski equation with flow and to a formal dependence of the effective diffusion constant on the anisotropoc flow field. However, a closed-form expression for the nucleation rate was not reported, because this requires solving the singularly-perturbed Smoluchowski equation with shear~\cite{Dhont}. 
On the whole, it is very difficult to rationalize all these very different outcomes, and apparently contradictory evidences, in the absence of a unifying, microscopic and analytical, description of the microscopic mechanism by which crystal nuclei form in sheared supercooled liquids.

Here we propose a microscopic analytical theory to provide a microscopic mechanism of the process, and possibly new insights into the qualitative physics. 
Deriving a fully quantitative theory of nucleation in shear flows is clearly impossible, as it is in fact already for the simpler nucleation without shear. However, we show
below that our new theory predicts qualitative behaviour for the nucleation rate in good agreement with recent simulation and experimental data. More importantly, it suggests a fundamental mechanism for the process, which was hitherto missing in the literature. 

In our derivation, we take a very fundamental approach, and start from the microscopic level of the master equation describing the nucleus formation by addition of atoms/molecules to sub-critical clusters. At this level, we account for the \textit{nonequilibrium} effect of shear flow on the transport of atoms/molecules to the cluster by using an analytical solution to the governing Smoluchowski diffusion-advection equation.
Further, we account for the effect of shear-induced mechanical deformation of the cluster.
Our fully analytical theory allows one to disentangle the different contributions of shear to the nucleation rate, and to predict the nucleation rate as a function of shear rate, and of other important physical and material parameters (e.g. molecular size, elasticity of the new phase cluster, viscosity of the melt etc.). The nucleation rate displays a pronounced maximum as a function of shear rate which we are able to explain qualitatively, for the first time, in terms of the competition between flow-induced advective  transport to the cluster and mechanical straining of the cluster.

\section{Derivation}
\subsection{Becker-Doering master equation for cluster growth}
Let us start by considering a fluid of diffusing particles (which could be atoms, molecules or colloidal particles), mutually interacting with an arbitrary intermolecular or interatomic interaction potential. The particles which constitute the supercooled liquid can aggregate by forming clusters (sub-critical nuclei) of different sizes. We follow here the original approach and notation of Zel'dovich~\cite{Zeldovich}. 
\begin{figure}
\centering
\includegraphics[width=0.62\columnwidth]{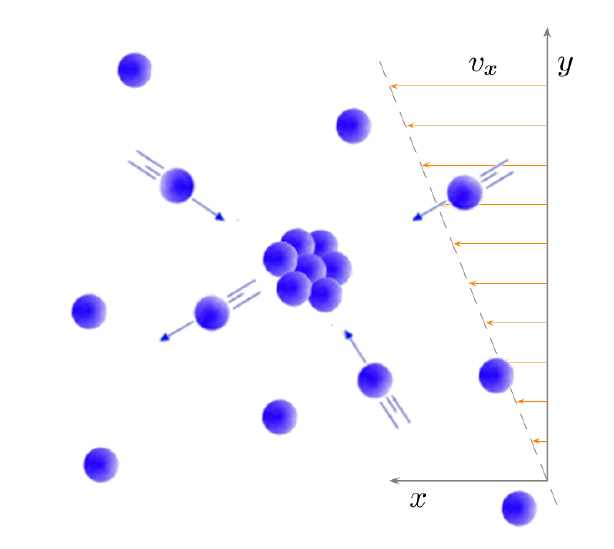}
\caption{Schematic of the cluster assembling process via microscopic single-particle addition and dissociation processes, in the presence of shear flow. The arrows schematically represent the flow velocity streamlines in a linear flow field. The flow velocity field at any point, in Cartesian components, is given by $v_{x}=\dot{\gamma}y$, where $\dot{\gamma}$ is the applied shear rate.
}
\label{fig:1}
\end{figure}

We let the coordinate $R$ be the radius of the spherical cluster, the growth thereof is described as a motion along the $R$-axis. The growth takes place in discrete jumps of length $\lambda$, i.e the radius variation due to the addition of a particle.
Since all cluster sizes are discretely distributed, the allowed sizes define a set of nodes  along the $R$-axis at distance $\lambda$ from each other. Nodes can be labelled with a discrete index $n$ expressing the number of particles forming the nucleus $n$.
We call the probabilities of a jump to the right (particle addition to the cluster)  or to the left (particle loss),  $q_+(n)$ and $q_-(n)$, respectively. \\
Therefore assuming the probability of the cooperative acquisition or loss of two or more monomers to be negligible, the variation in the probability density of nuclei $Z(n)$ at the $n^{\text{th}}$ node may be expressed by the following Becker-Doering equation,
    \begin{align}
	\label{Master1}
	\begin{split}
	\frac{\partial Z(n,t)}{\partial t}=& -Z(n,t)[q_+(n) + q_-(n)]+ \\
	&Z(n-1,t)q_+(n-1) + Z(n+1,t)q_-(n+1).
	\end{split}
	\end{align}
Denoting by $b(n)$ the equilibrium number of nuclei of size $n$, the principle of detailed balance gives: 
    \begin{align}
       \label{balance}
	\begin{split}
	&b(n)q_+(n)=b(n+1)q_-(n+1)\\
	&b(n-1)q_+(n-1)=b(n)q_-(n)
	\end{split}
	\end{align}
	We can thus eliminate from Eq.\eqref{Master1} all the dissociation rates $q_-$,  and denote the remaining rate $q_+$  simply by $q$. This leads to:
	    \begin{align}
		\label{Master2}
		\begin{split}
		\frac{\partial Z(n,t)}{\partial t}&=q(n)b(n)\left[ \frac{Z(n+1,t)}{b(n+1)} 		- \frac{Z(n,t)}{b(n)} \right]- \\
		& q(n-1)b(n-1) \left[ \frac{Z(n,t)}{b(n)} - \frac{Z(n-1,t)}{b(n-1)} 					\right].
		\end{split}
		\end{align}
We can turn to a continuous distribution with density $Z(R)$, by setting $Z(n)=\lambda Z(R)$, $Z(n+1)=\lambda Z(R+ \lambda)$, and so on. Assuming $\lambda$ to be a small constant number, we expand each term in Eq.\eqref{Master2} in a power series of $\lambda$; confining ourselves to the first non-vanishing term, one obtains the following diffusion equation in cluster-size space ~\cite{Zeldovich}
	    \begin{equation}
		\label{Fick1}
 		\frac{\partial Z}{\partial t}= \frac{\partial}{\partial R} 		  				\left(\lambda ^2qb 		\frac{\partial}{\partial R}\left( \frac{Z}{b} \right) 			\right)	 
 		=\frac{\partial}{\partial R} 		  				\left(Db 					\frac{\partial}{\partial R}\left( \frac{Z}{b} \right) 							\right)			  			
		\end{equation}
where the quantity $D=\lambda ^2q$ plays the role of a diffusion coefficient for the stochastic evolution in the space of cluster sizes~\cite{Zeldovich,Debenedetti}.
This is a crucial quantity which contains the microscopic physics of molecular transport towards the cluster, and thus it includes the effect of shear flow on the nucleus growth.

\subsection{Meaning of detailed-balance condition within classical nucleation theory}
Above and in the following, we apply the Zeldovich formulation classical nucleation theory (CNT), and our original contribution lies in the specification of the microscopic transport rate of a molecule to the cluster in shear flows, and in the derivation of the modified nucleation energy barrier to account for shear. Both these contributions are derived in the following sections and will be implemented within the Zeldovich framework for CNT subsequently. Here we would like to briefly discuss and contextualize the above derivation of a diffusive Fokker-Planck equation which is standard in classical CNT but raises some questions when applied to sheared systems.
For example, a fundamental question could be raised here about the validity or applicability of the detailed-balance condition within the Becker-Doering master equation in the context of nonequilibrium driven systems. 

More precisely, the detailed balance condition, in Eq.\eqref{balance}, is assumed in the one-dimentional nucleus size-space, but not in the phase-space of positions and momenta of the molecules, and returns an equilibrium state of zero current, $J(R,t)=0$, associated with the equilibrium distribution $b(R)\propto \exp[-F(R)/k_{B}T]$ . \\
Due to the shape of the free energy function in the nucleus size space, which shows a barrier for a critical size and becomes negative for larger sizes ( Fig.\ref{energiagrafico}), this solution would give a very large number of large nuclei after the barrier, and cannot predict the kinetic development of the nucleation process.
Thus, in the spirit of Zeldovich CNT we will solve Eq.\eqref{Fick1}  for a non equilibrium steady state $Z_{st}$ under the assumption $Z_{st}/b \sim 1$ before the barrier and $Z_{st}/b \sim 0$ after the critical size. Thus, within the same approach, we assume the ratio between the attachment and detachment rate in the equilibrium state to be the same also in the stationary nonequilibrium state. 

It is evident that nucleation, both with and without shear, is therefore always a nonequilibrium process which is accompanied by a non-zero flux ($J=const$). Detailed balance is just an initial condition which is useful to determine the dissociation rate (which is very difficult to quantify otherwise) as function of the association rate, and to eliminate it from the kinetic equations. 

Importantly, the existence of the initial quasi-equilibrium state in which $Z_{st}$ is very close to equilibrium distribution is justified, following Zeldovich theory, by the fact that the energy barrier is so steep that the initial probability of finding a cluster of critical size is extremely low, and hence $Z_{st}$ can be assumed to be thermalized and close to the Boltzmann form $b(R) \propto \exp[-F(R)/k_{B}T]$, but only until the barrier.

In other words, the system is initially localized (in the energy landscape) just below the steep energy barrier for nucleation, in a sort of "bound state" from which the escape process is so slow due to the high barrier that even if the distribution was not Boltzmann-like from the beginning, a stationary distribution, with features stated above, will have been established a long time before an appreciable number of clusters have escaped over the barrier. This is the same assumption underlying the derivation of Kramers' escape rate of a Brownian particle over a steep energy barrier ~\cite{Kramers}.

There is nothing obvious which forbids assuming a similar scenario for nucleation in shear flow as well, provided that, also in this case, the energy barrier for nucleation in size-space is also large and steep. In fact, as we will show below, the energy barrier with shear is even larger than in the absence of shear, which makes the above considerations even more reasonable and even more applicable for sheared systems compared to static systems.  

These arguments thus provide the justification for using detailed balance, within the Zeldovich assumption of initial quasi-equilibrium, to determine the dissociation rate in the microscopic derivation of nucleation theory in shear flows.

\subsection{Diffusion coefficient in cluster size space with shear}
In order to evaluate the diffusion coefficient in Eq.\eqref{Fick1} we have to estimate the probability $q$ that a single particle of radius $a$ joins a cluster of radius $R$  in the presence of shear flow.
The first obvious consideration concerns the diffusion coefficient of Brownian molecules in a shear flow. Due to the anisotropic geometry of shear, the effective diffusion influenced by shear becomes also anisotropic. For example, a formal expression for the diffusion coefficient in shear flow as a function of the flow field has been derived within the framework of mesoscopic nonequilibrium thermodynamics~\cite{Rubi}. Within the microscopic framework of the Smoluchowski equation with shear, this anisotropicity appears in the probability distribution function of particles in space which is the solution to the governing equation of motion. 
Within this approach, the aim is to calculate the rate of collision between a molecule and a cluster; a necessary step towards this aim is the evaluation of the flux over the cluster, which is a spherical isotropic integral of the probability distribution function of finding the molecule at a given distance from the cluster.  

In this way, the rate of a single-particle attachment to the cluster can be estimated by solving the Smoluchowski equation with shear or diffusion-advection equation~\cite{Levich, Dhont}, which governs the collision rate between the cluster and the single particle in the presence of: (i) the mutual Brownian diffusion of cluster and particle; (ii) the intermolecular interaction field between cluster and particle; (iii) the applied shear flow.

Let us consider spherical coordinates centred on the nucleus of radius $R$, and $c(r)$ the monomer concentration (or probability distribution function) averaged over angular coordinates $\left( \theta, \phi \right) $ in the spherical frame centred on the cluster. As shown in details in Appendix \ref{Smol_appendix}, $c(r)$ is the solution to the radial component of the two-body Smoluchowski diffusion-advection equation and defines the probability of finding a single molecule at a radial distance $r$ from a cluster. The radial component of the equation can be written as~\cite{Zaccone}:
\begin{equation}
		\label{smoluc1}
		\frac{1}{r^2}\frac{d}{dr} r^2 \left[  D \beta \left( \frac{dU}{dr} - B v_{r,\text{eff}}\right)  c  + Dr^2 \frac{d c } {dr} \right]=0
\end{equation}
where $U$ is the intermolecular interaction (which may also account for many-body correlations in an effective way, e.g. if one takes the potential of mean force) between the particle and the cluster. $D=D_a + D_R= k_{B}T (a+R)/6\pi \eta aR $ is the mutual diffusion coefficient with $\eta$ the solvent viscosity, $B=6\pi \eta aR / (a+R)$ is the hydrodynamic drag. 
$v_{r,\text{eff}}$ is the effective radial component of the relative velocity between the cluster and the particle due to the shear flow (see its definition in Appendix B).

In particular, $v_{r,\text{eff}}\neq0$ and is given by the standard radial component of the relative velocity, as given for simple shear flows~\cite{Russell}, only in those sectors of the solid angle where $v_{r}<0$. These are the sectors of solid angle where the flow brings the particles towards each other. In those sectors where, instead, the two particles are pushed away from each other and $v_{r}>0$, we take $v_{r,\text{eff}}=0$. The motivation for this simplification is that, upon taking the total inward flux, only those sectors of solid angle contribute to the inward flux where the flow brings the particles towards each other, whereas those sectors where the two particles are pushed away from each other by the shear field do not contribute to the inward flux. In this way, the anisotropic character of the flow is fully accounted for by the theory.

Importantly, this is not an uncontrolled simplification, but a necessary step for calculating the rate. This is reflected in the fact that different numerical values of the rate are given by this theory for different flow geometries (e.g. axysimmetric extensional flow, shear flow, sink flow, uniform flow etc), as discussed more in detail in previous work~\cite{Zaccone}. For example, within this approach, the value of the rate would be maximum for a sink radial flow~\cite{Zaccone08}, where the cluster is located at the sink point for the streamlines, since this is the most isotropic flow field, whereas the value is clearly much smaller for strongly anisotropic flow fields like shear.
One of course could think of solving the fully anisotropic Smoluchowski partial differential equation instead of its effective radial component (which is an ordinary differential equation), but this cannot be done analytically. It is instead possible to define the effective radial component of the flow field as done in~\cite{Zaccone}, which accounts for the anisotropicity of the flow, and use this within the Smoluchowski equation, Eq.(5), to calculate the collision rate which is a spherically averaged quantity by definition.   

It is also necessary to emphasize that the Smoluchowski equation with shear ensures that the anisotropic dynamics of Brownian particles is correctly described. A manifestation of this fact is that the \textit{local} collision rate according to Eq.(5) is anistropic and does depend on the angular orientation in the solid angle, while the \textit{total} inward flux is independent (by construction, being an integrated quantity) of the angular orientations. Another manifestation of the anisotropic dynamics predicted by the Smoluchowski equation with shear becomes evident if one transforms the Smoluchowski equation into its associated Langevin equation with shear. The latter, in turn, can be used to determine the mean squared displacement at steady-state as a function of time, in a standard way~\cite{Zwanzig}. The coefficient in this relation is an \textit{effective diffusion coefficient} which is manifestly anisotropic~\cite{Rubi}. Hence, the anisotropic diffusion is rather an outcome of Eq.(5), not an input to it. 

In dimensionless form, Eq.\eqref{smoluc1} becomes:
\begin{equation}
		\resizebox{.9\hsize}{!}{$
		\frac{1}{Pe(x+1)^2}\frac{d}{dx} (x+1)^2 \left[  \left(\frac{d \tilde{U}}{dx} 				-4Pe \tilde{v}_{r,\text{eff}} \right) c  + \frac{d  c }{dx} \right]=0
		$}
\end{equation}
where $x= \left[r/(R+a)\right] -1$, $Pe=\dot{\gamma}(a+R)^2/\left[4(D_a + D_R)\right]$ with $\dot{\gamma}$ the shear rate, and where we have introduced the non-dimensionalised potential and velocity, $\tilde{U}= \beta U$ and 
$\tilde{v}_{r,\text{eff}}=v_{r,\text{eff}}/\dot{\gamma}(R+a)$, respectively. 
We can set the boundary conditions for the collision problem:
\begin{align}
\label{boundary_condition}
	\begin{split}
		c  =0 &\quad\text{for  } x=0,\\
		c  = c_0 &\quad\text{for  } x=\delta/(R+a).
	\end{split}
\end{align}
where $c_0$ is the bulk density of molecules in the supercooled melt, and $\delta$ is the boundary-layer thickness which is defined below. 

As is known from many previous studies~\cite{Dhont}, the above Smoluchowski equation with shear is \textit{singularly-perturbed}, and presents a boundary-layer structure. In simple words, this means that no matter how small the Peclet number is, the equation cannot be solved by a simple perturbative expansion in $Pe$. This problem arises because the small parameter (the shear rate $\dot{\gamma}$ or the Peclet number, $Pe$) multiplies the relative radial velocity in the above differential equation in the term $-4Pe \tilde{v}_{r,\text{eff}}$, which, for all linear shear flows, diverges as $\dot{\gamma}r\rightarrow\infty$ in the far field limit, 
$r\rightarrow\infty$. 
The singular perturbation character means that no matter how small $Pe$ is, the flow term in the Smoluchowski equation is always going to diverge right at the far field limit where a boundary condition is required to integrate the differential equation. 
As a consequence, there exists an "outer layer", at $r>\delta$, where the shear advection term is always overwhelming compared to the other terms in the equation. 

Conversely, for sufficiently small separations, there exists an "inner layer" where all contributions (diffusion, potential, shear) are important. The width $\delta$ of the boundary layer thus separates the inner from the outer layer. Using the method of matched-asymptotics~\cite{Barenblatt}, one can thus develop a perturbative expansion in $1/Pe$ in the outer layer where the shear term dominates, and a different expansion in $Pe$ for the inner layer where shear does not dominate over the other terms. The two expansions can then be matched at the boundary layer $\delta$ to recover the full approximate solution over the entire domain of $r$. 

Since we are interested in determining the collision rate, we only need the inner layer solution, but we also need to know the location of the boundary layer because we need to take the surface integral of the concentration profile. It is found~\cite{Dhont2,Zaccone} that $\delta \propto Pe^{-1/2}$. Physically, this means that at very high Peclet numbers where the shear dominates over Brownian motion, the boundary layer is shifted towards very small separations, and the inner layer eventually shrinks to zero 
($\delta\rightarrow 0$) in the limit of $Pe\rightarrow\infty$. In this limit, the diffusive term in the equation can be dropped and the dynamics is entirely controlled by the flow advection~\cite{Dhont,Dhont2,Zaccone}. 

Anyway our interest is not directed towards this latter extreme case, because we want to limit our study to the range of small shear rate.
Thus the solution for the concentration profile inside the inner layer can be built here upon integrating the dimensionless equation, with boundary conditions Eq.\eqref{boundary_condition}, up to the non-dimensionalized boundary layer width $\frac{\delta}{R+a}$,
\begin{align}
	\begin{split}
		 c(r)  & = {\text{exp} \int_{\frac{\delta}{R+a}}^{x} 					dx\left(- \frac{d\tilde{U}}{dx}+ 4Pe \tilde{v}_{r,\text{eff}}\right) }\\
		& \times \Bigg[ c_0 + \frac{\Phi_0}{4 \pi (R+a)(D_a + D_R)}\int_{\frac{\delta}				{R+a}}^{x} \frac{dx}{(x+1)^2}\\
		& \times \text{exp}\int_{\frac{\delta}{R+a}}^{x}dx\left(\frac{d\tilde{U}}{dx} 		- 4Pe 				\tilde{v}_{r,\text{eff}}\right)\Bigg]
	\end{split}
\end{align}
where $\Phi_0$ is the inward flux of molecules colliding onto the cluster surface at $x= 0$ ~\cite{Zaccone}.
Solving for $\Phi_0$ we obtain
\begin{equation}
		 \Phi_0= \frac{8\pi (R+a)(D_a+D_R)c_0}{2 \int_{0}^{\frac{\delta}							{R+a}}\frac{dx}{(x+1)^2} \text{exp}\left[\int_{\frac{\delta}						{R+a}}^{x}dx\left(\frac{d\tilde{U}}{dx} - 4Pe \tilde{v}_{r,\text{eff}}\right)\right]}.
\end{equation} 
The change in single-particle concentration per unit time can be obtained upon multiplying the flux by the concentration of nuclei $c_R$, so that the kinetic equation for the rate reads as
\begin{equation}
        \label{secondord}
		\frac{dc_0}{dt}= -\Phi_0c_R= -\frac{8\pi(R+a)(D_a+D_R)}{W_\Phi} c_0c_R,
\end{equation}
where we have defined
\begin{equation}
        \resizebox{.85\hsize}{!}{$
		W_\Phi= 2 \int_{0}^{\frac{\delta}{R+a}} \frac{dx}{(x+1)^2} 							\text{exp}\left[ \int_{\frac{\delta}{R+a}}^{x} dx\left(\frac{d\tilde{U}}{dx} - 4Pe 				\tilde{v}_{r,\text{eff}}\right) \right].
		$}
\end{equation}
The Eq.\eqref{secondord} outlines the analogue of a second-order chemical reaction with the reaction rate given by
\begin{equation}
		\label{rate}
		q=\frac{4\pi (R+a)(D_a+D_R)c_0}{\int_{0}^{\frac{\delta}{R+a}} \frac{dx}				{(x+1)^2} 							\text{exp}\left[ 							\int_{\frac{\delta}{R+a}}^{x} dx \left(\frac{d\tilde{U}}{dx} - 4Pe 						\tilde{v}_{r,\text{eff}}\right) \right]}.
\end{equation}
We note that in the limit of $Pe=0$ and $U=0$, the rate $q$ correctly reduces to the diffusion-limited rate of a second-order chemical reaction~\cite{Nitzan}: $q=4\pi (D_a + D_R)(R+a)c_0$.
\subsection{Nucleus free energy under shear}
In classical nucleation theory~\cite{Debenedetti}, the free energy of a nucleus is the
sum of an enthalpy term which is proportional to the volume of the nucleus, and an interface term which is proportional to its surface. Taking these two contributes into account gives the standard free energy of the nucleus in the form~\cite{Debenedetti}
\begin{equation}
		F(R)=-\frac{4}{3}\pi R^3\frac{|\Delta \mu |}{v'} + 4\pi R^2 \nu,
\end{equation}
where $\Delta \mu <0$ is the difference in the chemical potential between the new phase (e.g. the crystal) and the metastable phase (e.g. the liquid), $v'$ is the volume of one particle, and $\nu$ is the surface tension.

As is well known, for small $R$ values the surface energy term dominates, the free energy thus increases with increasing $R$, while for large $R$ values the bulk enthalpy term dominates, and the free energy starts to decrease with increasing nucleus size $R$. Therefore, there is a range of $R$ where a free energy barrier arises with a maximum located at the critical value $R^*$, which defines the critical nucleus.
 Since our interest is in studying a system with shear, we have to consider also a further contribution $F_s$ in the free energy expression due to the nucleus deformation caused by the shear. If we consider a small elastic deformation due to the shear stress transmitted by the surrounding fluid motion onto the nucleus, the additional free energy contribution $F_s$ reads ~\cite{Landau, Slezov},
 \begin{equation}
 \frac{F_s}{V}=\frac{1}{2}\sigma _{ik}u_{ik}
\end{equation}  
where $V$ is the nucleus volume while $\sigma _{ik}$ and $u_{ik}$ represent the elastic stress and the symmetric component elastic of the strain tensor, respectively.\\

\begin{figure}
\includegraphics[width=0.50\textwidth]{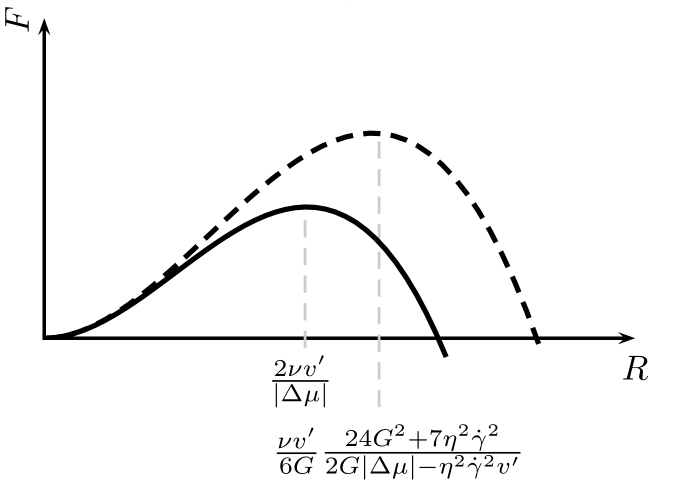}
\caption{Free energy as a function of the nucleus radius $R$. The continuous solid line is the dependence without shear, while the dashed line is the free energy in presence of shear as given by Eq. 17.\label{energiagrafico}}
\end{figure}

For a laminar shear flow, the hydrodynamic stress tensor in the fluid is given by ~\cite{Ogden}:
\begin{equation}
		\vec{\vec{\sigma}} =
 		\begin{pmatrix}
 		 0 & \eta \dot{\gamma} & 0 \\
  		\eta \dot{\gamma} & 0 & 0 \\
  		0 & 0 & 0\\
  		 \end{pmatrix}
\end{equation}
where $\eta$ is the fluid viscosity and $\dot{\gamma}$ is the shear rate.
We assume that the elastic stress acting upon the nucleus is equal to the average hydrodynamic stress in the sheared fluid. Upon considering that the elastic stress on the nucleus is given by  $\sigma _{xy}=2Gu_{xy}$ ~\cite{Landau}, where $G$ is the shear modulus of the nucleus, the elastic strain to which the nucleus is subjected is given by
\begin{equation}
\label{straintensor}
\vec{\vec{u}} =
 		\begin{pmatrix}
 		 0 & \frac{\eta \dot{\gamma}}{2G} & 0 \\
  		\frac{\eta \dot{\gamma}}{2G} & 0 & 0 \\
  		0 & 0 & 0\\
  		 \end{pmatrix}.
\end{equation} 
Then the total free energy of the nucleus becomes
\begin{equation}
\label{freenergy}
F=-\frac{4}{3}\pi R^3\frac{|\Delta \mu |}{v'} + 4\pi R^2 \nu \left( 1 + \frac{7}{24}\frac{\eta ^2  \dot{\gamma}^2 }{G^2}\right) + \frac{1}{2}\frac{\eta ^2 \dot{\gamma}^2}{G}\frac{4}{3}\pi R^3,
\end{equation}
which is depicted in Fig.\ref{energiagrafico}. 
In this derivation we accounted for the shape deformation of the spherical nucleus into ellipsoid due to shear stress. This effect results in a correction term to the nucleus surface which derivation is shown in more details in the Appendix \ref{ellips_appendix}. Without this correction, the surface term in the free energy of the nucleus would be simply $4\pi R^2 \nu$, as for a sphere. 
It is important to note that the shear-induced deformation doesn't  affect the volume of the nucleus which remains constant as expected for simple shear deformations which are volume-preserving.\\

However, the correction term for the nucleus surface change upon deformation into an ellipsoid, is negligible for a large class of systems where the deviation from spherical shape is small, and for some of those it has been indeed observed that this change happens at substantial values of shear rates, e.g. $\sim10-100 s^{-1}$ for polymers~\cite{Ryan}.
The deformation of the nucleus into an ellipsoid, furthermore, does not depend uniquely on the applied shear rate $\dot{\gamma}$, but it has to depend necessarily also on the shear rigidity of the nucleus. For example, if the nucleus were infinitely rigid, it would forever retain its spherical shape also at extremely high shear rates.
Since  the surface of the deformed ellipsoidal nucleus increases as $\Delta S \sim (\eta\dot{\gamma}/G)^{2}$, this relation gives the limit for which the nucleus can be described as spherical as long as $(\eta\dot{\gamma}/G)^{2}\ll 1$ is satisfied. 

\subsection{Analytical expression for the nucleation rate with shear}
The nucleation rate can be estimated using Kramers' escape rate theory for the crossing rate of the energy barrier~\cite{Kramers,Zwanzig}, when the growth of the clusters is governed by the Smoluchowski equation~\cite{Zeldovich}.
Starting from Eq.\eqref{Fick1}, and assuming the equilibrium distribution to have the Boltzmann form
\begin{equation}
		\label{MaxBoltz}
		b(R) \sim e^{- \frac{F (R)}{k_BT}},
\end{equation}
the associated current in cluster-size space is given by
\begin{equation}
\label{current}
J=-De^{-\frac{F (R)}{k_BT}}\frac{\partial}{\partial R} \left(Z e^{\frac{F (R)}{k_BT}}\right),
\end{equation}
where we recall that $D=\lambda ^2 q$ with $q$ given by the Eq.\eqref{rate} and $\lambda=a$ is the radius of a single particle.
At steady-state (i.e $J=\text{const}$) upon integrating Eq.\eqref{current} on both sides between $R_0\equiv0$ and a point $R_B$ located sufficiently far away beyond the barrier, and $R_B \text{ such that } F(R_B)\ll F(R_0)$, we obtain
\begin{align}
\begin{split}
J&= -\frac{Z^{st}(R)e^{\frac{F(R)}{k_BT}}\mid ^{R_B}_{R_0}}{\int _{R_0}^{R_B} dR\frac{1}{D} e^{\frac{F(R)}{k_BT}}} \simeq \frac{Z^{st}(R_0)e^{\frac{F(R_0)}{k_BT}}}{\int _{R_0}^{R_B} dR\frac{1}{D} e^{\frac{F(R)}{k_BT}}}\\
&\simeq\frac{Z^{st}(R_0)}{\int _{R_0}^{R_B} dR \frac{1}{D} e^{\frac{F(R)}{k_BT}}}
\end{split}
\end{align}
where $Z^{st}$ is the stationary distribution of cluster sizes.

The integral in the denominator is dominated by the exponential near the barrier, so neglecting the dependence of $D$ on $R$ (as discussed in~\cite{Zeldovich} and \cite{Frenkel}), expanding $F(R)$ in a second-order Taylor series around the maximum, and extending the limits in the integration domain to infinity, we find
\begin{align}
\begin{split}
J &\simeq\frac{Z^{st}(R_0)}{\frac{1}{D(R^*)}\int _{R_0}^{R_B} dR e^{\frac{F(R^*) + \frac{1}{2}F''(R^*)(R-R^*)^2}{k_BT}}}\\
&\simeq \frac{Z^{st}(R_0)D(R^*)}{\sqrt{\frac{2\pi k_BT}{-F''(R^*)}}e^{\frac{F(R^*)}{k_BT}}}.
\end{split}
\end{align} 
Near $R=0$ we can approximate the stationary distribution $Z^{st}(R)$ with the equilibrium Boltzmann distribution ~\cite{Zeldovich}, $b(R)$, which in $R_0$ is given by
\begin{equation}
Z^{st}(R_0)=b(R_0)=\frac{N_{tot}e^{-\frac{F(R=0)}{k_BT}}}{\int_{R_0}^{R^*} dR e^{-\frac{F(R)}{k_BT}}}
\end{equation}
where $N_{tot}$ is the total number of particles in the metastable state, before the barrier.

The integral in the denominator is dominated by the exponential near the origin, hence upon expanding $F(R)$ in a second-order Taylor series around $R_0$ and recalling that $F(R_0)=0$, we obtain
\begin{align}
\begin{split}
b(R_0)&\simeq \frac{N_{tot}}{\int_{R_0}^{R^*} dR e^{-\frac{1}{2}\frac{F''(R_0)(R-R_0)^2}{k_BT}}}\\
&\simeq \frac{2N_{tot}}{ \sqrt{\frac{2 \pi k_BT}{F''(R_0)}}}.
\end{split}
\end{align}
In the last step, we have extended the upper limit in the integration domain to infinity since contributions to the integral past the barrier are negligible~\cite{Kramers}.
Finally, the nucleation rate is defined by $K_N\equiv\frac{J}{N_{tot}}$ and reads
\begin{align}
\begin{split}
\label{nucleationrate}
K_N&= \frac{D(R^*)}{ \pi k_B T}\sqrt{-F''(R^*)F''(R_0)}e^{-\frac{F(R^*)}{k_BT}}\\
&=\frac{(8 \nu + \frac{7\dot{\gamma}^2\eta ^2 \nu}{3G^2}) D(R^*)}{k_BT}e^{-\frac{F(R^*)}{k_BT}}
\end{split}
\end{align}
where in the second line we considered $ F''(R_0)=-F''(R^*) =8 \pi \nu + \frac{7\pi \dot{\gamma}^2\eta ^2 \nu}{3G^2} $, obtained from Eq. \eqref{freenergy} with  
\[R^*=\frac{\nu v'}{6G}\frac{24G^2 + 7\eta ^2 \dot{\gamma}^2  }{2G\mid \Delta \mu \mid- \eta^2 \dot{\gamma}^2 v'}.\]

The free energy barrier at the critical nucleus size $R^*$ is found to be
\begin{equation}
\label{crit.freeenergy}
F(R^*)=\frac{\nu v'^2\nu^3 \pi}{648 G^4}\frac{(24G^2 + 7\eta ^2 \dot{\gamma}^2  )^3}{(2G\mid \Delta \mu \mid- \eta^2 \dot{\gamma}^2 v')^2}
\end{equation}
which increases as $\dot{\gamma}$ increases, as illustrated in Fig. \ref{energiagrafico}. 

In particular, upon Taylor-expanding this expression around $\dot{\gamma}=0$, we find that the first non-vanishing term in $\dot{\gamma}$ is the quadratic term, i.e. $F(R^*)\sim\dot{\gamma}^{2}$.
The effect of the shear rate on the total free energy is to increase the height of the barrier $F(R^{*})$, where $R^{*}$ is the critical nucleus size, and to shift its position to a higher $R$ value, thus slowing down the nucleation process. Therefore, our Eq.\eqref{nucleationrate}-\eqref{crit.freeenergy} explain the quadratic increase of the nucleation energy barrier as a function of shear rate,  which has been observed in numerical simulations in the past~\cite{Blaak}.

From a physical point of view, the quadratic increase of the free energy barrier for nucleation with the shear rate is explained within our framework in terms of the increased elastic energy of the nucleus imparted by the elastic straining due to hydrodynamic flow stress. It cannot be excluded that for certain systems the increase of nucleation barrier due to the increase of elastic energy may abruptly culminate with the breakup of fragmentation of the nucleus~\cite{Valeriani}, as the mechanical yielding of the nucleus may be achieved at high enough shear rates, a possibility discussed for example by Onuki~\cite{Onuki}, which certainly plays an important role for aggregating colloidal phases~\cite{Zaccone2013}.

It is also important to note that, within our theory, a critical value of shear rate $\dot{\gamma ^*}$ exists, for which the denominator in Eq.(\ref{crit.freeenergy}) goes to zero, causing the nucleation rate to vanish. This situation corresponds to the extreme case where not even the smallest infinitesimal nuclei would be mechanically stable under such large flow stresses, and nucleation is thus suppressed completely by the \textit{mechanical} instability of the new phase under the imposed shear stress. 

\subsection{Prefactor of the nucleation rate expression}
We can explicitly evaluate the prefactor in front of the exponential in Eq.\eqref{nucleationrate}\\
\begin{equation}
K_N ^0=\frac{(8 \nu + \frac{7\dot{\gamma}^2\eta ^2 \nu}{3G^2}) D(R^*)}{k_BT}.
\end{equation}
In the expression for the diffusion coefficient in size-space $D(R)$ we substitute the rate $q$ from Eq.\eqref{rate} and we get:
\begin{equation}
\label{prefattorecompleto}
K_N ^0=\frac{(8 \nu + \frac{7\dot{\gamma}^2\eta ^2 \nu}{3G^2}) 4a^2 \pi (R^*+a)(D_a+D_{R^*})c_0}{k_BT \int_{0}^{\frac{\delta}{(R+a)}} \frac{dx}				{(x+1)^2} 							\text{exp}\left[ 							\int_{\frac{\delta}{(R+a)}}^{x} dx \left( \frac{dU}{dx} - 4Pe 						\tilde{v}_{r,\text{eff}}\right)\right]}.
\end{equation}
In the presence of an intermolecular or interatomic interaction potential, with a range $\xi$, between the particles, it was previously established by means of scaling arguments and in comparison with full numerical simulations of the Smoluchowski diffusion-advection equation~\cite{Zaccone}, that $\delta/(R+a)\simeq ((R+a)Pe/\xi)^{-1/2}$.
In hard-sphere (HS) systems, the range of the bare pair-interaction is zero, by definition. However, the relevant interaction which causes the particles to stick onto a cluster is not the two-body pair potential, but rather the potential of mean force which features an attractive part with a finite range $\xi$, as discussed below.
%
%
%

\subsection{Calculation of the crystallization rate in sheared hard-sphere colloid systems}
Colloidal HS systems have been studied intensively both experimentally and computationally, as model systems to understand complex many-body dynamics and phase transitions. In the HS phase diagram, the controlling parameter is the volume fraction $\phi$ occupied by the colloidal particles. For example, HS systems undergo a first-order transition from liquid to crystal at the freezing packing fraction $\phi=0.54$, which is the analogue of the freezing temperature of atomic and molecular systems.  
Colloidal HS liquids at $\phi>0.54$ are therefore metastable and nucleation processes take place leading to the formation of the new crystal phase~\cite{Gasser}.

In HS systems, the bare two-body pair-potential is an infinitely steep wall and has zero range. However, at high particle density such as in the metastable regime 
$\phi>0.54$, many-body correlations lead to a potential of mean force which features a pronounced effective attraction between two particles. The potential of mean force is defined by $V_{mf}/k_B T=-\ln g(r)$, where $g(r)$ is the radial distribution function. The effective attraction between two nearest-neighbour particles arises due to the osmotic pressure, exerted by all the other particles, which remains unbalanced in the gap between the two particles~\cite{Dhont,Zaccone2012}.
Therefore, $V_{mf}$ cannot be confused with the simple two-body pair potential (which is just a hard wall here) because it crucially accounts for collective processes that are responsible for the cohesion of the crystal. 

This entropic effective attraction is what drives the attachment of a particle to a cluster or nucleus of the crystal nucleus, and is the consequence of many-body effects.  
The effective attraction has been calculated using different approaches, and it features an energy minimum of the order of $8-10 k_{B}T$ with a range $\xi\approx1.5\sigma=3a$, where $\sigma$ is the hard-sphere diameter~\cite{Schweizer}. 
Here, for our illustrative calculation, we assume that the potential of mean force is what governs the effective attraction between a particle freely moving in the supercooled liquid phase and a particle protruding on the cluster surface. 
Very schematically, we model the attraction as a ramp potential with an energy minimum of $-8 k_B T$ and range $\xi=3a$,
\begin{equation}
U=V_{mf}= \left[\frac{8k_BT}{3a}\left(r-(R-a)\right) - \frac{8k_BT}{3a}3a\right]\theta(R+2a-r)
\end{equation}
where $\theta$ is the Heaviside function. 

The qualitative behaviour of the denominator in Eq.\eqref{prefattorecompleto}, for a simple shear velocity field~\cite{Zaccone}:  $\tilde{v}_{r,\text{eff}}=- 1/3 \pi (x+1)$, as a function of Peclet number, can be easily estimated numerically and decreases as the Peclet number increases. 
Further, in the numerator the dependence of $D_a + D_{R^*}$ upon $\dot{\gamma}$ can be neglected in comparison with the dependence of $R^*$ on $\dot{\gamma}$ and the explicit dependence on $ \dot{\gamma}^2$ . Hence the prefactor $K_{N}^{0}$ of the nucleation rate displays an increasing trend with the shear rate.

In the expression of the nucleation rate, Eq.\eqref{nucleationrate}, two opposite contributions brought by the shear are present, in the prefactor (Eq.\eqref{prefattorecompleto}) and inside the exponential factor (Eq.\eqref{crit.freeenergy}), respectively. In fact, while the prefactor increases with the shear rate due to the enhancement of advective-diffusive transport towards the nucleus, the exponential factor decreases upon increasing the shear rate due to the increased elastic energy of the nucleus which increases the nucleation barrier. 
As a consequence of this competition (prefactor increasing with $\dot{\gamma}$, exponential factor decreasing with $\dot{\gamma}$), an overall non-monotonic dependence of the nucleation rate upon the shear rate, with a point of maximum, arises. 

\begin{table}
\centering
\caption{Parameters values for a colloidal suspension of PMMA spheres in a mixture of decahydronaphthalene and cyclohexylbromide. The nucleation rate obtained with these values is plotted in Fig.\ref{PlotRateColloids}. The parameter values are taken from~\cite{Gasser}, with the exception of the viscosity which has been tuned in our calculation to recover the experimentally measured nucleation rate in the absence of shear. \label{tabella}}
\setlength{\extrarowheight}{0.2cm}
\setlength{\tabcolsep}{0.5 cm}
\begin{tabular*}{\hsize}{@{\extracolsep{\fill}}ccc}
   Parameter & Value & Units  \cr
  \hline

   $\Delta \mu $ & $5.25 \times 10^{-22}$ &$J$    \cr

   $\eta$& $1.8 \times 10^{-1}$  &$Pa \cdot s$   \cr

   $\nu $& $6.87 \times 10^{-11}$ &$N/m $ \cr
 
   $G $ & $1.6 \times 10^{-3}$& $Pa$ \cr
   $c_0 $ & $6 \times 10^{16}$&$1/m^{3}$ \cr
   $k_BT $ & $4 \times 10^{-21}$&$J$  \cr
   
   \noalign{\vskip 0.1cm}
   \hline
\end{tabular*}
\end{table}

We calculated the nucleation rate on the example of the crystallization of a HS colloidal suspension of poly(methyl methacrylate) (PMMA) spheres in a mixture of decahydronaphthalene and cyclohexylbromide. If not stated otherwise, parameters values, reported in Tab.\ref{tabella}, are taken from Ref.~\cite{Gasser}. The viscosity $\eta \approx 1.8 \times 10^{-1}$ $Pa \cdot s$ is estimated by the calibration of our theoretical prediction of nucleation rate in absence of shear, with the experimental results of  Ref.~\cite{Gasser}. It is important to note that the chemical potential difference between crystal and liquid 
$\Delta \mu$ is in general a function of the control parameter which for colloids is volume fraction $\phi$ (it would be the temperature in atomic systems), and the same applies to the viscosity. These parameters therefore introduce a dependency on the supersaturation which here we do not consider explicitly and we focus on a fixed quench into the metastable regime. 

\begin{figure}[t]
\centering
\includegraphics[width=0.50\textwidth]{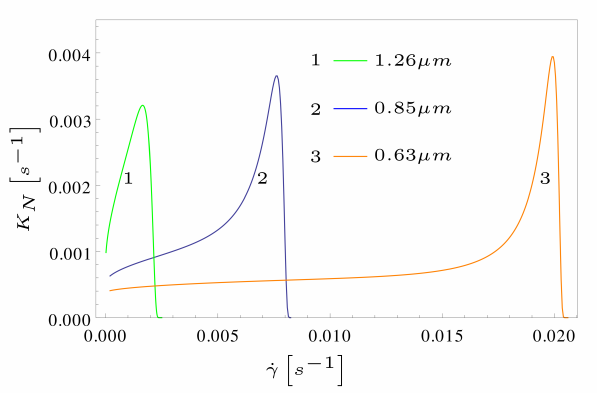}
\caption{Nucleation rate of BCC colloidal crystals as a function of the shear rate $\dot{\gamma}$, plotted using the parameters values reported in Tab.\ref{tabella}. Different curves are obtained for different values of the particles radius.\label{PlotRateColloids}}
\end{figure}

On theoretical grounds~\cite{Alexander}, the first phase formed near the melting line is the BCC crystal phase, although the stable phase is the FCC crystal. For our illustrative calculations, we assume the BCC structure, although of course the calculation can be done for any crystal structure using the Born-Huang theory of elastic constants of crystals. Therefore, the shear modulus $G$ is estimated using the standard Born-Huang formula for BCC crystals~\cite{Born} $G=\frac{2}{3}\frac{\kappa}{l}$ assuming that only nearest-neighbours matter. Hence, using $\kappa\approx 10k_{B}T/l^2$, we estimate $G=10 \cdot\frac{2}{3} k_BT/l^3$ where $l \approx 2a$ is the crystal lattice constant.
The nucleation rate with shear flow for this system was calculated using Eq.\eqref{nucleationrate}, and is plotted in Fig.\ref{PlotRateColloids} for selected values of the physical parameters. \\

As shown in Fig.3, the nucleation rate increases with the shear rate until it reaches a maximum value for an optimal value of shear rate $\dot{\gamma}^{*}$.
The three curves in Fig.\ref{PlotRateColloids} are obtained upon varying colloid size: as the latter decreases, a significant shift of the optimal shear value takes place, while the peak amplitude remains almost constant. The physical origin of this effect is partly controlled by the nucleus elasticity: smaller particles make stronger nuclei and the increase of elastic energy becomes important at comparatively higher $\dot{\gamma}$, while at lower $\dot{\gamma}$ the nucleation rate is comparatively lower because the advective-diffusive transport towards the nucleus is slower with smaller particles (which have smaller $Pe$ numbers).
If the particle-size effect was solely controlled by the nucleus elasticity, we would expect a dependence of the nucleation rate peak on the shear rate as to the cubic power, because the elastic modulus scales as $~k_{B}T/a^{3}$. The dependence is however somewhat stronger, to the fourth power, because of the size effect due to molecular transport.  

\section{Simulations and experiments: from colloids to atomic systems}
Recent simulation results ~\cite{Cerdà, Mokshin, Speck}, showed the existence of a peak in the nucleation rate with a characteristic non-monotonic dependence of nucleation rate on the shear rate. The nucleation rate is sped up at low shear rates up to the peak value, after which it decreases with further increasing shear rate. This qualitative behaviour was not explained by any clear or simple microscopic mechanism thus far, and to our knowledge the theory presented here provides the first mechanistic explanation of this effect in terms of the competition between advective enhancement of molecular transport to the nucleus and increased energy barrier due to straining. \\
The same qualitative behaviour has been reported recently for the nucleation kinetics of amyloid aggregation in shear flow~\cite{Hirsa}, whereby the nucleation rate extracted based on a Finke-Watzky model features a maximum as a function of the shear rate.

While this qualitative agreement is certainly encouraging, a more quantitative comparison with either simulations or experiments is still out of reach. The main issue here is the unavailability of physical parameters such as e.g. the viscosity, the nucleus' shear modulus, its surface energy or its volume-energy term, which are not provided in previous studies. Also, whenever they were measured, these parameter values are often model-dependent or based on assumptions which are foreign to our theory. For example, the nucleation energy barrier in~\cite{Mokshin} was extracted based on the assumption of an effective temperature which is unnecessary and redundant in our framework where the effect of shear is described at the microscopic level of molecular motion and cluster growth. The estimate of the energy barrier provided by those authors takes into account also the microscopic effect of shear on molecular transport and using it in our framework would lead to counting this effect twice since we already account for it in a different way. \\
In future simulations or experimental studies, these parameters need be estimated independently of any model assumption for the specific systems under study, to allow a more quantitative comparison with predictions of the theory presented here. Of course this is a very challenging task for which no solution is yet in sight. 

Calculations similar in spirit to the illustrative predictions for colloidal crystallization presented above could be done, in principle, for atomic systems as well, such as silicon~\cite{Kerrache}, or more complex metallic melts~\cite{Osuji}. 
It is expected that the peak of shear rate in those systems be found at much higher shear rates (in fact of order $1 s^{-1}$ as reported recently for metallic melts~\cite{Osuji}), due to the much smaller size of the building blocks (atoms instead of colloidal particles). However, extrapolating our theory for colloids presented above by just replacing the colloid size with an atomic size, would predict that shear effects become important only for enormously high shear rates. This unreasonable outcome for atomic systems is due to the fact that the theory for colloids cannot be extrapolated to atomic systems without, at the same time, adjusting the other physical parameters such as the viscosity and the shear modulus which are many orders of magnitude larger in atomic systems. This effect can be understood by considering the important role played by the viscosity in modulating the effect of shear on the atomic motion. The Peclet number increases linearly with the viscosity, but our prefactor in the nucleation rate depends exponentially on the Peclet number. The viscosity in atomic systems is 10 orders of magnitude larger than in colloidal systems (it increases strongly nonlinearly with decreasing the building block size), which makes the effect of shear remain important at accessible shear rates in atomic systems, thus compensating the effect of decreasing the size building block size. 

Finally, an other word of caution should be spent about comparing theoretical predictions to simulation data. Currently used protocols vary from Langevin dynamics where shear flow is treated as an external force in the equation of motion, to nonequilibrium molecular dynamics employing microscopic equations of motion such as the DOLLS or the SLLOD equation of motion~\cite{Evans}. Each of these methods has a number of idiosyncrasies (e.g. the SLLOD equation of motion cannot be derived from a Hamiltonian, whereas the DOLLS yields erroneous results at moderate shear rate) which makes any comparison with analytical theory a highly non-trivial business. Another possible issue of discrepancy in such a comparison comes from the role of boundary conditions and boundary effects in simulations. It is clear that, whenever periodic boundary conditions (e.g. Lees-Edwards) are not employed, particles moving close to the driven boundary experience non-trivial entropic effects while their affinity to the wall of the simulation box introduces another source of important additional effects, which are absent in analytical theories that work in the homogeneous, thermodynamic limit. On the other hand, also the use of Lees-Edwards periodic boundary conditions is not free from arbitrary assumptions (starting from the choice of way particles are re-inserted into the system as they cross a boundary), and different implementations are available~\cite{Pagonabarraga}. 
\section{Conclusions}
Nucleation phenomena in liquids are always occurring under some external perturbation, especially in industrial settings and in biological systems.
Shear flow is the paradigmatic external drive to approximate mechanical perturbations on otherwise quiescent, equilibrium systems. 
We started off from the basic (Becker-Doering) master equation for the nucleus self-assembly by molecular transport-driven attachment and detachment of molecules to and from the nucleus cluster.

Using a matched-asymptotics approximation of the singularly-perturbed diffusion-advection dynamics (Smoluchowski equation with shear), within the Zel'dovich reduction of the Becker-Doering equation to a Fokker-Planck equation in cluster-size space, we were able to estimate the effect of shear flow on the growth rate of nuclei and to derive a closed-form expression for the nucleation rate. The latter step is achieved using Kramers' method to evaluate the rate of crossing the nucleation barrier. \\
Further to the effect on the transport rate of molecules towards the nucleus, the shear flow also affects the energetics of the nuclei. 
The shear flow imparts shear stress on the nucleus which reacts elastically, and this increases its energy. For the case of crystallization, the contribution of shear stresses to the energy barrier for nucleation is always positive (energy barrier increases due to strain), and can be estimated in good approximation using the Born-Huang theory of crystal elasticity for different lattice structures.

This framework delivers an analytical theory of crystallization kinetics in shear. The main outcome of the theory is the non-monotonic dependence of the nucleation rate on the shear-rate. At low shear rates, the nucleation rate increases with shear rate because of the increase in advective transport towards the nucleus. As the shear rate increases further, the increase in the elastic energy of the strained nucleus becomes more and more important, which increases the nucleation energy barrier. The competition between these two opposite contributions (flow advection and shear-induced strain energy, respectively) is responsible for the appearance of a maximum in the nucleation rate. Past the maximum, the nucleation rate starts to decrease upon further increasing the shear rate as the controlling effect becomes the increase in strain energy leading to higher barriers inside the exponential Arrhenius factor. 
This framework opens up the possibility of understanding nucleation in flowing systems, with widespread applications, from shear-induced crystallization in metallic melts, to protein crystallization under physiological conditions. 
Also, it may help the rational design of experimental systems for the direct verification of the laws predicted by our theory.\\

\begin{acknowledgments}
F. M. is supported by a DFG Fellowship through the Graduate School of Quantitative Biosciences Munich (QBM). Discussions with Professor Daan Frenkel are gratefully acknowledged.
\end{acknowledgments}
\appendix
\section{Smoluchowski equation with shear for the orientation-averaged concentration field}
\label{Smol_appendix}
In order to derive Eq.(\ref{smoluc1}) is convenient to start from the full Smoluchowski equation for a sheared system:

\begin{equation} \label{GrindEQ__5_} 
\vec{\nabla}\cdot\left[\beta D\left(-\vec{\nabla} U+b\vec{v}\right)-D\vec{\nabla} \right]c=0 
\end{equation} 
 with the associate current:
\begin{equation} \label{GrindEQ__6_} 
\vec{J}=\left[\beta D\left(-\vec{\nabla} U+b\vec{v}\right)-D\vec{\nabla} \right]c 
\end{equation}

The incoming flux of particles on a spherical surface is given by:
\begin{align}
\begin{split}
\Phi&=\oint \vec{J} \cdot  \hat{\vec{n}}dS =\oint  \left[-D \vec{\nabla} c-\beta D\left(\vec{\nabla} U-b\vec{v}\right)c\right] \cdot \hat{\vec{n}} dS \\
&=\oint  D \left(\beta \frac{dU}{dr}c - \beta b v_r c + \frac{dc}{dr} \right) r^2 \sin\theta d\theta d\phi \\
 &=4\pi D r^2 \left[ \beta \left( \frac{dU}{dr}\langle c\rangle - b\langle v_r c \rangle \right) +  \frac{d\langle c \rangle}{dr}\right] 
\end{split}
\end{align}
where $\vec{\hat{n}}$ is the unit vector directed inwardly. Since we are interested in the net inward flux of particles, we can run the angular integration only on those angles such that the radial component of the velocity field is negative (which corresponds to the two particles being advected into each other by the flow).
It is thus convenient to define an effective radial velocity which depends on the angular orientation as:
\begin{equation}
v_{r,\text{eff}}  = \begin{cases}v_r & \text{if }  v_r <0  \\ 
0 & \text{otherwise} .
\end{cases}
\end{equation}

Under this assumption the inward flux becomes:
\begin{equation}
\Phi=4\pi D r^2 \left[ \beta \left( \frac{dU}{dr}\langle c\rangle - b\langle v_{r,\text{eff}} c \rangle \right) +  \frac{d\langle c \rangle}{dr}\right] 
\end{equation}
and supposing that convection is not overwhelming Brownian motion we can also assume a weak correlation between the concentration profile and the velocity field:
\begin{equation}
\langle v_r(\vec{r})c(\vec{r})\rangle \simeq  \langle v_r(\vec{r}) \rangle \langle c(\vec{r})\rangle
\end{equation}
which allows us to obtain an analytical expression for the flux as:
\begin{equation}
\Phi=4\pi D r^2 \left[ \beta \left( \frac{dU}{dr} - B\langle v_{r,\text{eff}}  \rangle \right) +  \frac{d}{dr}\right] \langle c\rangle.
\end{equation}
It is possible to show that the same result can be obtained starting directly from the following Smoluchowski equation: 
\begin{equation} \label{smol_eff} 
\vec{\nabla}\cdot\left[\beta D\left(-\vec{\nabla} U+B\vec{v}_{r ,\rm{eff}}\right)-D\vec{\nabla} \right]\langle c \rangle=0 
\end{equation} 
where we defined the effective (inwardly directed) velocity field as $\vec{v}_{r ,\rm{eff}}= [\langle v_r ^-\rangle , 0,0]^{\text{T}}$.
Writing Eq.(\ref{smol_eff}) for the radial coordinate as appropriate for determining the flux, and setting $\langle c \rangle$ for economy of notation, we recover Eq.(\ref{smoluc1}) of the main text.

\section{Shear-induced deformation of the nucleus into an ellipsoid}
\label{ellips_appendix}
The action of a simple shear flow described by the fluid-flow strain tensor for simple shear flow $\vec{\vec{s}}$. It is important to note that this is different from the elastic strain tensor in Eq.(16), which describes the elastic contribution to the free energy and has to be necessarily symmetric and cannot include rotational components (which are associated with dissipation).  The fluid strain tensor $\vec{\vec{s}}$, instead, must include also the rotational component 
and gives rise to an affine deformation $\vec{X'}= \vec{\vec{T}} \vec{X}$ where $\vec{X}$ is a generic point in 3D space and:
\begin{equation}
\vec{\vec{T}}= \mathbb{1} + \vec{\vec{s}}=
\begin{pmatrix}
 		 1 & \alpha & 0 \\
  		0 & 1 & 0 \\
  		0 & 0 & 1\\
  		 \end{pmatrix}.
\end{equation}
where we defined $\alpha =\frac{\eta \dot{\gamma}}{G} $. \\

We are now interested in observing how this deformation modifies the surface and volume of a spherical object in the limit of small $\alpha$. 
For the sake of simplicity let us consider a unitary sphere described by the equation:

\begin{equation}
\vec{X}^T\mathbb{1}\vec{X}=1.
\end{equation}
Under the action of $T$ the equation becomes that of an ellipsoid:
\begin{equation}
\vec{X'}^TT^{-T}T^{-1}\vec{X'}=1.
\end{equation}

Writing the quantity $T^{-T}T^{-1}$ in diagonal form, gives
\begin{equation}
\begin{pmatrix}
 		1&0 & 0 \\
  		0 & \frac{2+\alpha ^2 - \alpha \sqrt{(4+\alpha ^2)}}{2} & 0 \\
  		0 & 0 &\frac{2+\alpha ^2 +\alpha \sqrt{(4+\alpha ^2)}}{2}\\
  		 \end{pmatrix}.
\end{equation}

The eigenvalues of this matrix represent the lengths of semi-axes of the ellipsoid with equation:
\begin{equation}
\frac{x^2}{a^2} +\frac{y^2}{b^2} + \frac{z^2}{c^2}=1.
\end{equation}
Using the Legendre's approximated expression~\cite{Legendre} to calculate the surface of an ellipsoid:
\begin{equation}
S=4\pi ab\left( \frac{2}{3} + \frac{c^2b^2 + c^2a^2}{6a^2b^2}  \right)
\end{equation}
and Taylor expanding around $\alpha=0$ we obtain:
\begin{equation}
\label{surf_corr}
S = 4 \pi +\frac{7 \pi  \alpha ^2}{6}+O[\alpha ]^3
\end{equation}
 which gives a correction to a spherical surface of the second order in $\dot{\gamma}$.\\

\end{document}